# Scattering of a Single Plasmon by Three Non-equally Spaced Quantum Dots System Coupled to One-Dimensional Waveguide

Nam-Chol Kim,[†,*]  Myong-Chol Ko,[†]

[†]Department of Physics, Kim Il Sung University, Pyongyang, D P R Korea

**Abstract:** Scattering properties of a single plasmon interacting with three non-equally spaced quantum dots coupled to one-dimensional surface plasmonic waveguide is investigated theoretically via the real-space approach. It is demonstrated that the transmission and reflection of a single plasmon can be switched on or off by controlling the detuning and changing the interparticle distances between the quantum dots. By controlling the transition frequencies and interparticle distances of QDs, one can construct a half-transmitting mirror with three QDs system. We also showed that controlling the transition frequencies and interparticle distances of QDs results in the complete transmission peak near the zero detuning.

**Keywords:** Switching, Scattering, Plasmon, Waveguide

[*] Electronic mail: ryongnam10@yahoo.com

# 1. Introduction

The light - matter interaction has always been a fundamental topic in physics, and its most elementary level is the interaction between a single photon and a single emitter [1]. Controlling the scattering properties of a single photon has attracted particular interests for some fundamental investigations of photon-atom interaction and for its applications in quantum information[2]. Since photons are attracted and regarded as ideal carriers of quantum information, photons are naturally considered to replace electrons in future information technology[2]. Recently, theoretical idea of a single-photon transistor has also emerged[3]. Such a nonlinear device is essential to many emerging technologies, such as optical communication, optical quantum computer, and quantum information processing. Many theoretical[4-12] and experimental[13-15] works reported the photon scattering in different quantum systems. In the previous studies, the authors have mainly considered the scattering properties of a single photon interacting with one emitter and several emitters, and they mainly focus on the case where the quantum emitters are all equally spaced each other[16-19].

The coupling between metal nanowires and quantum emitters is important for controlling light-matter interactions[20]. The nanowire exhibits good confinement and guiding even when its radius is reduced well below the optical wavelength. In this limit, the effective Purcell factor $P \equiv \Gamma_{pl} / \Gamma'$ can exceed $10^3$ in realistic systems according to the theoretical results[21], where $\Gamma_{pl}$ is the spontaneous emission rate into the surface plasmons(photons) and $\Gamma'$ describes contributions from both emission into free space and non-radiative emission via ohmic losses in the conductor. Both experimental and theoretical investigations show that the emission properties of QDs can be significantly modified near the metallic nanostructures[22,23]. Recent investigations have further extended into the regime of interactions of quantum emitters such as QDs and propagating SPs[24]. The recent progress of the photonic crystal waveguide could be a step toward the goal of minimizing the radiation loss to the surrounding, as some theoretical studies have indicated that the high spontaneous emission can be achieved for systems with a quantum dot doped inside a photonic crystal waveguide. Motivated by these considerations, we investigate the scattering of a single plasmon interacting with three non-equally spaced QDs coupled to 1D surface plasmonic waveguide which is a metal nanowire [Fig. 1].

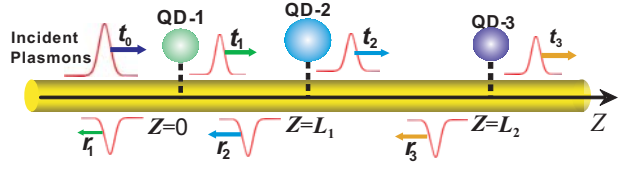

**FIG.1.** (Color online) Schematic diagram of a system consisting of a single plasmon and three non-equally spaced QDs coupled to a metal nanowire. $t_i$ and $r_i$ are the transmission and reflection amplitudes at the place $z_i$, respectively.

In this paper, we investigate theoretically the scattering properties of a single plasmon by non-equally spaced QDs coupled to a one-dimensional(1D) surface plasmatic waveguide, where the transition frequencies of the three QDs can be different with each other.

## 2. Theoretical model and dynamics equations

We investigate the scattering of a single plasmon interacting with three non-equally spaced QDs coupled to 1D surface plasmonic waveguide which is a metal nanowire as shown in Fig. 1.

Under the rotating wave approximation, the Hamiltonian of the system in real space is given by [4,5]

$$H = \sum_{j=1}^{3} \left[ (\omega_e^{(j)} - i\Gamma_j'/2)\sigma_{ee}^j + \omega_g^{(j)}\sigma_{gg}^j \right] + iv_g \int_{-\infty}^{\infty} dz \left[ a_l^+(z)\partial_z a_l(z) - a_r^+(z)\partial_z a_r(z) \right] \\ + \sum_{j=1}^{3} g_j \left\{ \left[ a_r^+(z_j) + a_l^+(z_j) \right]\sigma_{ge}^j + \left[ a_r(z_j) + a_l(z_j) \right]\sigma_{ge}^j \right\} \quad (1)$$

where $\omega_e^{(j)}$ and $\omega_g^{(j)}$ are the eigenfrequencies of the state $|g\rangle$ and $|e\rangle$ of the $j$ th QD, respectively, $\omega_k$ is the frequency of the propagating plasmon field with wavevector $k$ ($\omega_k = v_g|k|$). $\sigma_{ge}^j = |g\rangle_{jj}\langle e|$ ($\sigma_{eg}^j = |e\rangle_{jj}\langle g|$) is the lowing (raising) operators of the $j$ th QD, $a_r^+(z_j)$ ($a_l^+(z_j)$) is the bosonic operator creating a right-going (left-going) plasmon at position $z_j$ of the $j$ th QD. $v_g$ is the group velocity of the corresponding $\omega_k$, the non-Hermitian term in $H$ describes the decay of state $\omega_e^{(j)}$ at a rate $\Gamma_j'$ into all other possible channels. $g_j = (2\pi\hbar/\omega_k)^{1/2}\Omega_j \mathbf{D}_j \cdot \mathbf{e}_k$ is the coupling constant of the $j$ th QD with a single plasmon, $\Omega_j$ is the resonance energy of the $j$ th QD, $\mathbf{D}_j$ is the dipole moment of the $j$ th QD, $\mathbf{e}_k$ is the polarization unit vector of the surface plasmon[4,5].

Assuming that a single plasmon is incoming from the left with energy $E_k = \hbar\omega_k$, then the eigenstate of the system, defined by $H|\psi_k\rangle = E_k|\psi_k\rangle$, can be constructed in the form

$$|\psi_k\rangle = \int dz \left[\phi_{k,r}^+(z) a_r^+(z) + \phi_{k,l}^+(z) a_l^+(z)\right]|0, g\rangle + \sum_{j=1}^{3} e_k^{(j)}|0, e_j\rangle, \qquad (2)$$

where $|0, g\rangle$ denotes the vacuum state with zero plasmon and three QDs being unexcited, $|0, e_j\rangle$ denotes the vacuum field and only the $j$ th QD in the excited state and $e_k^{(j)}$ is the probability amplitude of the $j$th QD in the excited state. $\Phi_{k,r}^+(z)$ ($\Phi_{k,l}^+(z)$) is the wavefunction of a right-going (a left-going) plasmon at position $z$.

For a single plasmon incident from the left, the mode functions $\Phi_{k,r}^+(z)$ and $\Phi_{k,l}^+(z)$ can take the forms as followings, $\phi_{k,r}^+(z<0) = e^{ikz}$, $\phi_{k,r}^+(0<z<L_1) = t_1 e^{ik(z-L_1)}$, $\phi_{k,r}^+(L_1<z<L_1+L_2) = t_2 e^{ik(z-L_1-L_2)}$, $\phi_{k,r}^+(z>L_1+L_2) = t_3 e^{ik(z-2L_1-2L_2)}$, $\phi_{k,l}^+(z<0) = r_1 e^{-ikz}$, $\phi_{k,l}^+(0<z<L_1) = r_2 e^{-ik(z-L_1)}$, $\phi_{k,l}^+(L_1<z<L_1+L_2) = r_3 e^{-ik(z-L_1-L_2)}$ and $\phi_{k,l}^+(z>L_1+L_2) = 0$, respectively, where $L_1$ is the spacing between the first and the second QDs and $L_2$ between the second and the third QDs. Here $t_j$ and $r_j$ are the transmission and reflection amplitudes at the place $z_j$, respectively. By substituting Eq. (2) into $H|\psi_k\rangle = E_k|\psi_k\rangle$, we obtain a set of equations as: $r_2 e^{ikL_1} - r_1 - \frac{ig_1}{v_g} e_k^{(1)} = 0$,

$t_1 e^{-ikL_1} - t_0 + \frac{ig_1}{v_g} e_k^{(1)} = 0$, $\quad t_0 + r_1 - \frac{\Delta_1}{g_1} e_k^{(1)} = 0$, $\quad r_3 e^{ikL_2} - r_2 - \frac{ig_2}{v_g} e_k^{(2)} = 0$,

$t_2 e^{-ikL_2} - t_1 + \frac{ig_2}{v_g} e_k^{(2)} = 0$, $\quad t_1 + r_2 - \frac{\Delta_2}{g_2} e_k^{(2)} = 0$, $\quad r_4 e^{ik(L_1+L_2)} - r_3 - \frac{ig_3}{v_g} e_k^{(3)} = 0$,

$t_3 e^{-ik(L_1+L_2)} - t_2 + \frac{ig_3}{v_g} e_k^{(3)} = 0$, $\quad t_2 + r_3 + \frac{\Delta_3}{g_3} e_k^{(3)} = 0$, where $\omega_g^{(j)} = 0$, $\Omega_j = \omega_e^{(j)} - \omega_g^{(j)}$,

$\Delta_k^{(j)} \equiv \Omega_j - \omega_k$, $J_j \equiv g_j^2/v_g$, $\Delta_j \equiv i\Gamma_j'/2 - \Delta_k^{(j)}$, $(j=1, 2, 3)$. By taking the boundary conditions of the mode functions, i. e., $t_0 + r_1 = t_1 e^{-ikL_1} + r_2 e^{ikL_1}$, $t_1 + r_2 = t_2 e^{-ikL_2} + r_3 e^{ikL_2}$, $t_2 + r_3 = t_3 e^{-ik(L_1+L_2)} + r_4$, where $t_0 = 1$, $r_4 = 0$, into account, we obtain the transmission and the reflection amplitudes, respectively. By evaluating the transmission coefficient $T_3 \equiv$ $|t_3|^2$ and reflection coefficient $R_1 \equiv |r_1|^2$, we can find the scattering properties of a single plasmon in the long time limit.

### 3. Theoretical analysis and numerical results

First of all, we pay attention to the dependency of transmission spectra on the spacing between the QDs. Figure 2 shows typical transmission spectra versus $kL_2$ for the cases that all the three transition frequencies of the QDs are equal to each other, where we set the phases $kL_1$ as (a) $kL_1 = 1.1\pi$, (b) $kL_1 = 1.2\pi$ and (c) $kL_1 = 1.37\pi$, respectively. Now, we suppose that $J_i = J$ = const ($j$ = 1, 2, 3). As shown in Fig. 2(a), the transmission coefficient versus phase $kL_2$ exhibits periodic properties, the period of which is $kL_2=\pi$. In other words, the period is a half of the wavelength of incident plasmon calculated in terms of spacing $L_2$, where we took the relation $k = 2/\lambda$ into account. We also found that this oscillatory pattern of transmission spectra dose not depend on the spacing between the first QD and the second QD. Figure 2(b) shows that setting the spacing between the first QD and the second QD as $kL_1 = 1.2\pi$ results in the switching between the complete transmitting and half – transmitting by controlling the spacing between the second QD and the third QD. What is more interesting is that, whatever the spacing between the second QD and the third QD, $L_2$, is, setting the spacing $L_1$ as $kL_1 = 1.37\pi$ results in the constant transmission coefficient (≈0.84) [Fig. 2(c)]. The result implies that there is a possibility to build a body with such a constant transmission coefficient as needed, which can provide a new idea of designing quantum optical devices such as a filter or a switch in its single-plasmon level.

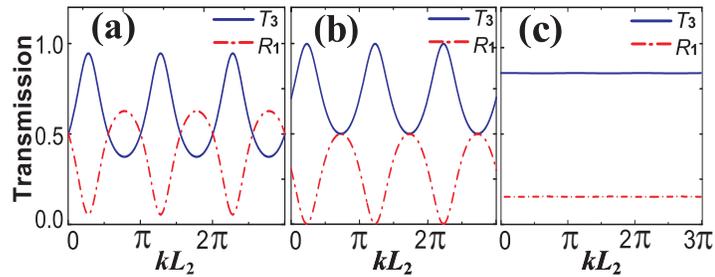

**Fig. 2.** (Color online) Transmission spectra of a propagating plasmon interacting with three non-equally spaced QDs ($\Delta_k^{(i)}$=0.000232$\omega_0$, $i$ = 1, 2, 3) versus $kL_2$, where the frequency is in units of $\omega_0 \equiv 2\pi v_g / L_1$ and $J = 10^{-4} \omega_0$: (a) $kL_1$=0.8$\pi$, (b) $kL_1$=1.37$\pi$, (c) $kL_1$=1.37$\pi$.

Figure 3 shows the transmission spectra versus $kL_2$ when the two QDs have the same detunings $0.000232\omega_0$ and the another has a detuning $0.0001\omega_0$. As shown in Fig. 3(a), the transmission spectra exhibit an oscillatory pattern with half of the wavelength of the propagating plasmon as a period in terms of spacing between $L_2$. From Figs. 2(b) and 2(c), we also found that the order of the QDs can influence on the scattering of the incident plasmon, which is quite different with the case of two QDs system [17]. Especially, we can construct a half-transmitting mirror with three QDs system by controlling the transition frequencies of QDs and setting the spacing $kL_1$ as $1.37\pi$, as shown in Fig. 3(c). From the result shown in Fig. 3, a half-transmitting mirror for a single incident plasmon can be built with three QDs by controlling the detunings and the spacings between QDs, which is valuable for the practical applications.

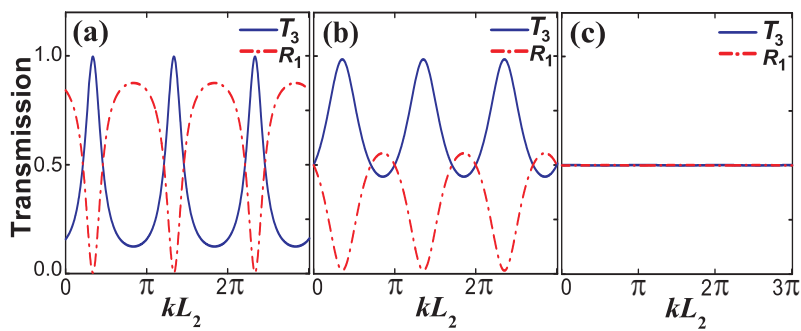

**Fig. 3.** (Color online) Transmission spectra of a propagating plasmon interacting with three non-equally spaced QDs versus $kL_2$, where $\omega_0 \equiv 2\pi v_g / L_1$ and $J = 10^{-4}\omega_0$: (a) $kL_1=0.8\pi$, $\Delta_k^{(1)} = \Delta_k^{(2)} = 0.000232\omega_0$, $\Delta_k^{(3)}=0.0001\omega_0$, (b) $kL_1=1.37\pi$, $\Delta_k^{(1)} = \Delta_k^{(3)} = 0.000232\omega_0$, $\Delta_k^{(2)}=0.0001\omega_0$, (c) $kL_1=1.37\pi$, $\Delta_k^{(1)} = \Delta_k^{(2)} = 0.000232\omega_0$, $\Delta_k^{(3)}=0.0001\omega_0$.

Now, we can also consider the transmission spectra versus the frequency of incident plasmon $\omega_k$. When all the three QDs have the same transition frequencies ($\Omega_1 = \Omega_2 = \Omega_3 = 1.012\Omega$), the transmission spectra are shown in Figs. 4(a) and 4(b). There appears a sharp complete transmission peak near the zero detuning when the spacings are equal to each other, $L_1 = L_2 = (0.1+0.5n)\lambda$, $n = 0, 1, 2, \cdots$ [Fig. 4(a)]. As is well known [17], a single plasmon switch has two states (on and off) for controlling the transmission of a single plasmon, in which the switch is off for $T_3=0$, and on for $T_3=1$. As shown in Fig. 4(a), the switch is off when the propagating plasmon is resonant with QDs and on when the propagating plasmon is detuned with QDs a little, which could be applied in desiging switches of single plasmons. When the spacings are different with each other, $L_1 \neq L_2$, there doesn't appear a complete transmission peak but a semi-permeable peak of transmission near the zero detuning [Fig. 4(b)], which is quite different with the case of equal spacings as shown in Fig. 4(a). Anyway, the results shown in Fig. 4 show that the spacing between QDs influence the scattering properties of an incident single plasmon greatly.

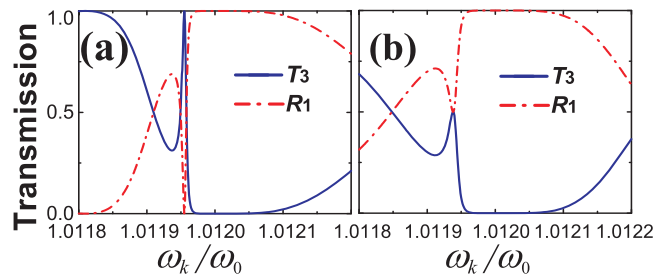

**Fig. 4.** (Color online) Transmission spectra of a single plasmon interacting with three QDs versus incident frequency of a single plasmon: (a) $kL_1=kL_2=0.2\pi$ (b) $kL_1=0.2\pi$, $kL_2=0.41\pi$, where $\Omega_1=\Omega_2=\Omega_3=1.0120\omega_0$, $\omega_0 \equiv 2\pi v_g / L_1$ and $J = 10^{-4}\omega_0$.

Figure 5 shows the transmission properties of a single plasmon interacting with non-equally spaced QDs having different transition frequencies. The solid line in Fig. 5 shows the transmission coefficient when the spacing $L_1$ is larger than the spacing $L_2$ ($kL_1=0.6\pi$, $kL_2=0.1\pi$). The dash-dot line in Fig. 5 shows the transmission coefficient when $L_1 < L_2$ ($kL_1=0.1\pi$, $kL_2=0.6\pi$). As we can see easily form Fig. 5, when $L_1= (0.3+0.5n)\,\lambda$, $L_2 = (0.05+0.5n)\,\lambda$ ($n = 0, 1, 2, \cdots$), where we took the relation $k =2 / \lambda$ into account, there appears a very sharp complete transmission peak near the frequency $\Omega_1=\Omega_2=1.0130\omega_0$, which is quite interesting. In contrast, there is no sharp complete transmission peak near the zero detuning and there appears a complete transmission peak at $(\Omega_1+\Omega_2) / 2$, when $L_1 < L_2$ ($kL_1=0.1\pi$, $kL_2=0.6\pi$). From Fig. 5, we found that the interparticle distances influence the scattering properties of single incident plasmon and doping QDs having the same transition frequencies result in the complete transmission peak near the zero detuning, which could find practical applications.

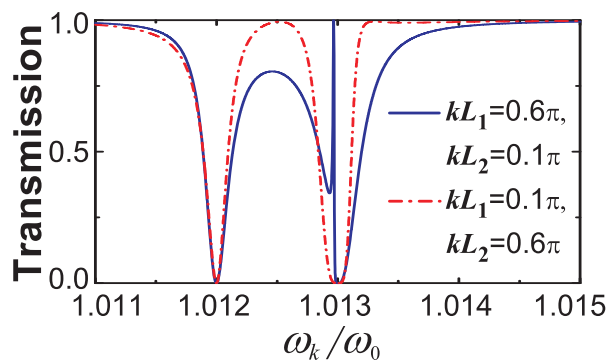

**Fig. 5.** (Color online) Transmission spectra of a single plasmon interacting with three QDs versus incident frequency of a single plasmon: We set $\Omega_1=\Omega_2=1.0130\omega_0$, $\Omega_3=1.0120\omega_0$, $\omega_0 \equiv 2\pi v_g / L_1$ and $J = 10^{-4}\omega_0$. The solid line(blue) displays the transmission spectra when $kL_1=0.6\pi$, $kL_2=0.1\pi$, and the dash-dot line(red) displays the transmission spectra when $kL_1=0.1\pi$, $kL_2=0.6\pi$.

## 4. Conclusions

In summary, we investigated theoretically the switching properties of a single plasmon interacting with three non-equally spaced QDs which are coupled one dimensional plasmatic waveguide. We showed that the transmission and reflection of a single plasmon can be switched on or off by dynamically tuning and changing the spacing between the QDs. Setting the interparticle distances properly results in the switching between the complete transmission and half – transmission. Especially, we can construct a half-transmitting mirror with three QDs system by controlling the transition frequencies and interparticle distances of QDs, which is valuable for the practical applications. We also showed that controlling the transition frequencies and interparticle distances results in the complete transmission peak near the zero detuning. Our results may find a variety of applications in the design of the quantum optical devices, such as nanomirrors and quantum switches, and in quantum information processing.

**Acknowledgments.** This work was supported by the National Program on Key Science Research and Key Project for Frontier Research on Quantum Information and Quantum Optics of Ministry of Education of D. P. R of Korea.